\documentclass[aps,prl,twocolumn,superscriptaddress]{revtex4-1}
\setcitestyle{super}

\usepackage{graphicx,epstopdf,siunitx,braket,xspace,amssymb,dsfont,bm,natbib}
\usepackage{todonotes}
\newcommand{\mc}[1]{\ensuremath{\mathcal{#1}}}
\newcommand{\mr}[1]{\ensuremath{\mathrm{#1}}}
\newcommand{\eps}[0]{\ensuremath{\epsilon}\xspace}
\newcommand{\fid}[0]{\ensuremath{\mc{F}}\xspace}

\newcommand{\leak}[0]{\ensuremath{\mc{L}}\xspace}

\newcommand{\sts}[0]{$\mathrm{S\mbox{-}T_0}$\xspace}

\newcommand{\op}[1]{\ensuremath{\hat{#1}}\xspace}

\newcommand{\reftab}[1]{Tab.\,\ref{#1}}

\DeclareRobustCommand{\reffig}[1]{Fig.\,\ref{#1}\,}

\newcommand{\m}[1]{%
    \IfEqCase{#1}{%
        {1}{\ensuremath{\op{\sigma}_{30}}}%
        {2}{\ensuremath{\op{\sigma}_{03}}}%
    }[\PackageError{m}{Undefined option to m: #1}{}]%
}%
\newcommand{\n}[1]{%
    \IfEqCase{#1}{%
        {qubits}{\ensuremath{n}\xspace}%
    }[\ensuremath{{n_{\mr{#1}}}}\xspace]%
}

\newcommand{\Tte}[0]{\ensuremath{T_2^{\mathrm{echo}}}\xspace}
\newcommand{\Tts}[0]{\ensuremath{T_2^*}\xspace}
\newcommand{\dbz}[0]{\ensuremath{\mr{\Delta}B_z}\xspace}
\newcommand{\je}[0]{\ensuremath{J(\epsilon)}\xspace}
\newcommand{\qS}[0]{\ensuremath{\Ket{\mathrm{S}}}\xspace}
\newcommand{\qT}[0]{\ensuremath{\Ket{\mathrm{T_0}}}\xspace}
\newcommand{\qDU}[0]{\ensuremath{\Ket{\downarrow\uparrow}}\xspace}
\newcommand{\qUD}[0]{\ensuremath{\Ket{\uparrow\downarrow}}\xspace}

\newcommand{\figlabel}[1]{\textbf{#1}}
\DeclareRobustCommand{\citenumber}[1]{Ref.\,\citenum{#1}\xspace}

\newcommand{\qZ}[0]{\ensuremath{\Ket{0}}\xspace}
\newcommand{\qO}[0]{\ensuremath{\Ket{1}}\xspace}
\newcommand{\jz}[0]{\ensuremath{J_0}\xspace}
\newcommand{\epsz}[0]{\ensuremath{\epsilon_0}\xspace}

\begin{document}

\title{Closed-loop control of a GaAs-based singlet-triplet spin qubit with 99.5\,\% gate fidelity and low leakage}

\author{Pascal Cerfontaine}
\email{pascal.cerfontaine@rwth-aachen.de}

\author{Tim Botzem}
\affiliation{These authors contributed equally to this work}
\affiliation{JARA-FIT Institute for Quantum Information, Forschungszentrum J\"ulich GmbH and RWTH Aachen University, 52074 Aachen, Germany}

\author{Julian Ritzmann}
\affiliation{Lehrstuhl f\"ur Angewandte Festk\"orperphysik, Ruhr-Universit\"at Bochum, D-44780 Bochum, Germany}

\author{Simon Sebastian Humpohl}
\affiliation{JARA-FIT Institute for Quantum Information, Forschungszentrum J\"ulich GmbH and RWTH Aachen University, 52074 Aachen, Germany}

\author{Arne Ludwig}
\affiliation{Lehrstuhl f\"ur Angewandte Festk\"orperphysik, Ruhr-Universit\"at Bochum, D-44780 Bochum, Germany}

\author{Dieter Schuh}
\affiliation{Institut f\"ur Experimentelle und Angewandte Physik, Universit\"at Regensburg, D-93040 Regensburg, Germany}

\author{Dominique Bougeard}
\affiliation{Institut f\"ur Experimentelle und Angewandte Physik, Universit\"at Regensburg, D-93040 Regensburg, Germany}

\author{Andreas D. Wieck}
\affiliation{Lehrstuhl f\"ur Angewandte Festk\"orperphysik, Ruhr-Universit\"at Bochum, D-44780 Bochum, Germany}

\author{Hendrik Bluhm}
\email{bluhm@physik.rwth-aachen.de}
\affiliation{JARA-FIT Institute for Quantum Information, Forschungszentrum J\"ulich GmbH and RWTH Aachen University, 52074 Aachen, Germany}
\date{August 18, 2020}


\maketitle
\textbf{Semiconductor spin qubits have recently seen major advances in coherence time and control fidelities, leading to a single-qubit performance that is on par with other leading qubit platforms. Most of this progress is based on microwave control of single spins in devices made of isotopically purified silicon. For controlling spins, the exchange interaction is an additional key ingredient which poses new challenges for high-fidelity control. Here, we demonstrate exchange-based single-qubit gates of two-electron spin qubits in GaAs double quantum dots. Using careful pulse optimization and closed-loop tuning, we achieve a randomized benchmarking fidelity of $\mathbf{(99.50 \pm 0.04)\,\%}$ and a leakage rate of $\mathbf{0.13\,\%}$ out of the computational subspace. These results open new perspectives for microwave-free control of singlet-triplet qubits in GaAs and other materials.}

\section{Introduction}
While semiconductor spin qubits have been pioneered with GaAs-based quantum dot devices \cite{Elzerman2004, Petta2005, Hanson2007a, Barthel2009, Bluhm2010, Nowack2011, Shulman2012}, the adoption of isotopically purified silicon to avoid decoherence from nuclear spins has led to coherence times approaching one second \cite{Veldhorst2014, Muhonen2014} and control fidelities above \SI{99.9}{\%} \cite{Yoneda2017, Muhonen2015, Yang2019a}, thus meeting the requirements for scalable quantum computing regarding single-qubit performance. These results are generally achieved with resonant microwave control of individual spins via electric or magnetic fields. An additional key requirement is to controllably couple multiple qubits. A natural and widespread approach is to use the exchange interaction between tunnel-coupled electron spins, which can also be used to manipulate qubits encoded in two or more electron spins \cite{DiVincenzo2000, Levy2002}. Advantages of this approach include a short gate duration and the avoidance of microwaves, which is a considerable simplification regarding power dissipation, complexity of control systems and addressability in the context of scaling to large qubit numbers. Exchange-based two-qubit gates of individual spins as well as two-electron single-qubit gates have reached fidelities up to about \SI{98}{\%} \cite{Huang2018a, Nichol2017} before our work. However, qubit control via the exchange interaction is also associated with certain challenges like the need for strong driving well beyond the rotating wave approximation, nonlinear coupling to control fields, a susceptibility to charge noise that scales with the interaction strength \cite{Dial2013}, and a high sensitivity to the detailed shape of baseband control pulses.

To address these difficulties, we numerically optimized control pulses for exchange-based single-qubit gates with fidelities approaching \SI{99.9}{\%} in previous work \cite{Cerfontaine2014}. Remaining inaccuracies in these optimized pulses can be removed by a closed-loop gate set calibration protocol (GSC), which allows the iterative tune-up of gates using experimental feedback \cite{Cerfontaine2014, Cerfontaine2019a}. In comparison to automated calibration of single-spin \cite{Yang2019a} and superconducting qubits \cite{Egger2014, Kelly2014}, GSC extracts tomographic information about all unitary degrees of freedom to improve convergence. In addition to recalibration of drifting parameters, this method is suited for tune-up of gates with initial infidelities larger than \SI{10}{\%} \cite{Cerfontaine2014}. Here, it allows us to optimize roughly an order of magnitude more parameters than before \cite{Yang2019a,Egger2014,Kelly2014} to fully leverage the degrees of freedom provided by our hardware. 

With this approach we achieve accurate control of GaAs-based singlet-triplet qubits encoded in two electron spins with a fidelity of \SI{99.50 \pm 0.04}{\%} (in contrast to a preliminary preprint \cite{Cerfontaine2016} of the present study we used a device with a more representative charge noise level compared to other groups \cite{Dial2013}). For comparison, state-of-the-art single-spin control in GaAs leads to a fidelity of \SI{96}{\%} \cite{Yoneda2014}. Furthermore, our result is in the range required by certain quantum error correction schemes \cite{Fowler2009,Rispler2020} and validates corresponding simulations \cite{Cerfontaine2014}, which also predict a similar performance for two-qubit control \cite{Cerfontaine2019}. A comparable fidelity of \SI{99.6}{\%} has also recently been demonstrated for singlet-triplet single-qubit gates in Si \cite{Takeda2019}. In addition, we demonstrate a low leakage rate of \SI{0.13}{\%} out of the subspace of valid qubit states, an important consideration for any qubit encoded in multiple spins \cite{Wallman2015}. A similar leakage rate of \SI{0.17}{\%} has been observed in an isotopically purified Si device using a three-spin encoding \cite{Andrews2018}. Besides addressing the challenges of exchange-based qubit control, these results also open new perspectives for GaAs-based devices.

\section{Results}
\subsection{Singlet-triplet qubit}
\begin{figure}
	\centering
	\includegraphics[width=7.5cm]{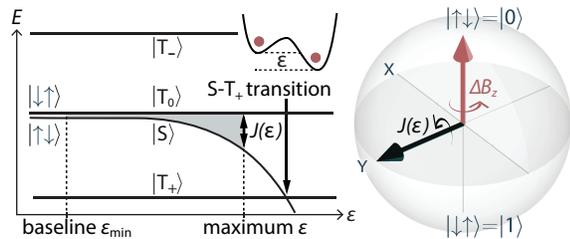}
	\caption{\textbf{$\mathbf{S-T_0}$ qubit energy diagram and Bloch sphere.} The eigenenergies change as a function of detuning \eps, which is used to control the exchange coupling \je. The \eps pulses presented in this work start and finish at a baseline $\eps_{\min}$ with low $J$ and pulse to higher values for short periods. The maximum amplitude is constrained to below the $\mathrm{S\mbox{-}T_+}$ anticrossing at large \eps. We choose the convention that \je points along the \textit{y}-axis of the Bloch sphere. For low \eps amplitudes, the qubit rotates about \dbz, the \textit{z}-axis of the Bloch sphere. Large amplitude \eps pulses rotate the qubit about the \textit{y}-axis and thus enable arbitrary single-qubit gates.}
\label{fig:conv}	
\end{figure}

The \sts spin qubit \cite{Levy2002} used in this work is illustrated in the middle of the figure below and can be described by the Hamiltonian $H = \frac{\hbar \je}{2} \sigma_x + \frac{\hbar \dbz}{2} \sigma_z$ in the $(\qUD = \qZ, \qDU = \qO)$ basis, where arrows denote electron spin up and down states. \je denotes the exchange splitting between the singlet $\qS = (\qUD-\qDU)/\sqrt{2}$ and $s_z=0$ triplet state $\qT = (\qUD+\qDU)/\sqrt{2}$, while \dbz is the magnetic field gradient across both dots from different nuclear spin polarizations \cite{Petta2005}. The remaining triplet states, $\ket{\mathrm{T_+}} = \ket{\uparrow\uparrow}$ and $\ket{\mathrm{T_-}} = \ket{\downarrow\downarrow}$, represent undesirable leakage states. \je is manipulated by the detuning \eps, the potential difference between both dots. We use standard state initialization and readout based on electron exchange with the lead, Pauli blockade and charge sensing (see methods). For single qubit operations, \eps is pulsed on a nanosecond timescale using an arbitrary waveform generator (AWG) whereas \dbz is typically stabilized at $2 \pi \cdot \SI{42.1 \pm 2.8}{MHz}$ by dynamic nuclear polarization (DNP) \cite{Bluhm2010}. The resulting dynamics are illustrated in \reffig{fig:conv}. In our simulations we use the experimentally motivated model $\je = J_0 \exp{(\eps/\eps_0)}$\cite{Dial2013} to capture the nonlinear relation between control voltage and exchange coupling. 

\begin{table}[b]
\small
  \centering  
  	\begin{ruledtabular}
    \begin{tabular}{rrcl}
    Sequences $U_i$ (right to left) & Parametrization & & $S_i$ \\
    \hline
    $\pi/2_x$ & $-2\phi$  & = &$S_1$\\
    $\pi/2_y$ & $-2\chi$  & = &$S_2$\\
    $\pi/2_y \circ \pi/2_x$ & $-n_y - n_z - v_x - v_z$  & = &$S_3$\\
    $\pi/2_x \circ \pi/2_y$ & $-n_y + n_z - v_x + v_z$  & = &$S_4$\\
    $\pi/2_x \circ \pi/2_x \circ \pi/2_x \circ \pi/2_y$ & $n_y + n_z + v_x - v_z$  & = &$S_5$\\
    $\pi/2_x \circ \pi/2_y \circ \pi/2_y \circ \pi/2_y$ & $n_y - n_z + v_x + v_z$  & = &$S_6$\\
    $\pi/2_x \circ \pi/2_x$ & $d_x$ & = & $S_7$\\
    $\pi/2_y \circ \pi/2_y$ & $d_y$ & = & $S_8$\\       
    \end{tabular}%
    \end{ruledtabular}   
  \caption{\textbf{Tomographic gate sequences.} To first order, the outcome of the measurement $\mr{Tr}(\sigma_z U_i \qZ\!\Bra{0} U_i^\dagger) = S_i$ depends linearly on the gates' rotation-angle errors $2\phi$ ($2\chi$), the axis-errors $n_y, n_z$ ($v_x, v_z$) and decoherence $d_x$ ($d_y$) of the $\pi/2_x$-gate ($\pi/2_y$-gate). Parametrization defined as in \citenumber{Cerfontaine2014} and \citenumber{Dobrovitski2010}.}   
  \label{tab:bs}%
\end{table}%

\begin{figure*}
	\centering
	\includegraphics[trim=0.6cm 0.0cm 0.4cm 0, clip, width=2.06\columnwidth]{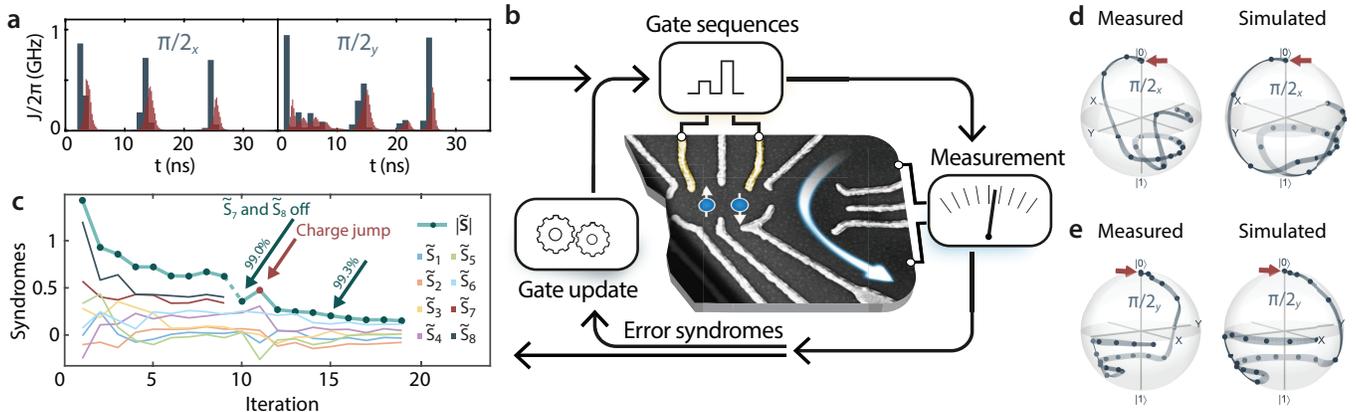}
	\caption{\textbf{Gate set calibration (GSC)} 
	\figlabel{a} Numerical pulse optimization based on a realistic but inaccurate qubit model provides initial optimal control pulses (blue) for \SI{36}{ns} long $\pi/2_x$ and $\pi/2_y$ gates. According to the model, the pulses shown in red are actually seen by the qubit. 	
	\figlabel{b} Next, these pulses are optimized on the experiment using closed-loop feedback. 8 error syndromes $\tilde{S_i}$ are extracted in each iteration by applying the gate sequences from \reftab{tab:bs}. In order to remove gate errors, the syndromes $\tilde{S_i}$ are minimized by adjusting the pulse segments' amplitudes $\eps_j^g$. 	
	\figlabel{c} Typically, GSC converges within 15 iterations and can recover from charge rearrangements in the quantum dot (indicated by a red dot, see Supplementary Note 16). Before iteration 1, the gate fidelity is typically so low that randomized benchmarking \cite{Magesan2011} (RB) can not be used to reliably extract the gate fidelity. This is remedied by scaling the pulses before the first iteration, leading to an average Clifford gate fidelity between \SI{63}{\%} and \SI{70}{\%}. In this specific calibration run, the feedback loop improved the fidelity of the gate set first to \SI{99.0 \pm 0.1}{\%}, then to \SI{99.3 \pm 0.1}{\%} (after disabling the decoherence syndromes $\tilde{S_7}$ and $\tilde{S_8}$) and eventually to \SI{99.50 \pm 0.04}{\%} (after adding a small correction of $0.05$ to $S_{\mathrm{M}}$). All of these fidelities are extracted using RB.	
	\figlabel{d-e} For a different gate set consisting of two \SI{24}{ns} long pulses, we performed self-consistent state tomography \cite{Takahashi2013}. After a few GSC iterations, the simulated Bloch sphere trajectories (right) can be reproduced in the experiment (left). A major portion of the remaining deviation can be attributed to concatenation errors with the measurement pulses, specifically when states following large $J$ pulses are determined.
}
\label{fig:main}	
\end{figure*}

\subsection{Numerical pulse optimization}
To experimentally implement accurate single qubit $\pi/2$ rotations around the $x$- and $y$-axis (denoted by $\pi/2_x$ and $\pi/2_y$), we use a control loop adapted from \citenumber{Cerfontaine2014} (illustrated in \reffig{fig:main}a-b) in conjunction with numerically optimized control pulses. To obtain a reasonably accurate system model for the numerical optimization procedure, we measure the step response of our electrical setup, \jz, \epsz and \dbz. In addition, we determine the coherence properties of the qubit to construct a noise model including quasistatic hyperfine noise, quasistatic charge noise and white charge noise. The details of the noise and control model are discussed further below and in Supplementary Notes 1 to 6. Next, we use this model to numerically optimize pulses consisting of $N_\mathrm{seg}$ piece-wise constant nominal detuning values $\eps_j, j = 1 \dots N_\mathrm{seg}$ to be programmed into the AWG with a segment duration of \SI{1}{ns}. The last four to five segments are set to the same baseline level $\eps_{\min}$ for all gates to minimize  errors arising from pulse transients of previous pulses. We choose $\eps_{\min}$ such that $J(\eps_{\min}) \ll \dbz$. Typical optimized pulse profiles $\eps_j^g, j = 1 \dots N_\mathrm{seg}$ for two gates $g = \pi/2_x$ and $g = \pi/2_y$ are shown in \reffig{fig:main}a. 

\subsection{Experimental gate set calibration}
Since our control model does not capture all effects to sufficient accuracy to directly achieve high-fidelity gates, these pulses need to be refined using experimental feedback. Hence, error information about the gate set is extracted in every iteration of our control loop. Standard quantum process tomography cannot be applied to extract this information as it requires well-calibrated gates, which are not available before completion of the control loop. We solve this bootstrap problem with a self-consistent method that extracts 8 error syndromes $S_i, i = 1 \ldots 8$ in each iteration \cite{Cerfontaine2014}. The first six syndromes are primarily related to over-rotation and off-axis errors while the remaining two syndromes are proxies for decoherence. A syndrome $S_i$ is measured by preparing \qZ, applying the corresponding sequence $U_i$ of gates from \reftab{tab:bs}, and determining the probability $p(\qZ)$ of obtaining the state \qZ by measuring the sequence $10^3$ to $10^4$ times. For perfect gates, the first six syndromes\cite{Dobrovitski2010} should yield $p(\qZ) = 0.5$, corresponding to $S_i = \left<\sigma_z\right> = 0$. The last two syndromes should yield $p(\qZ) = 0$ ($S_i = -1$). Deviations of $S_i$ from the expected values indicate decoherence and systematic errors in the gate set. To make our method less sensitive to state preparation and measurement (SPAM) errors, we also prepare and read out a completely mixed state with measurement result $S_{\mathrm{M}}$, and a triplet state \qT, which yields the measurement result $S_{\mathrm{T}}$ after correcting for the approximate contrast loss of the triplet preparation (see Supplementary Note 12). GSC then minimizes the norm of the modified error syndromes $\tilde{S_i} = S_i - S_{\mathrm{M}}$ for $i = 1 \ldots 6$ and $\tilde{S_i} = S_i - S_{\mathrm{T}}$ for $i \in \{7, 8 \}$.

For swift convergence, we start the control loop with numerically optimized pulses $\eps_j^g$ which theoretically implement the desired operations without systematic errors and with minimal decoherence by partially decoupling from slow noise (similar to dynamically corrected gates \cite{Khodjasteh2009a}). First, we scale these pulses by \SI{\pm 10}{\%} in \SI{2}{\%} increments and measure which scaling achieves the lowest $\tilde{S_i}$. GSC then optimizes the best pulses by minimizing $\tilde{S_i}$ with the Levenberg-Marquardt algorithm (LMA). In each LMA iteration, we use finite differences to experimentally estimate derivatives $d\tilde{S_i}/d\eps_j^g$, which are subsequently used to calculate updated pulse amplitudes $\eps_j^g$. Note that \dbz is not calibrated since the control via DNP is more involved than adjusting \eps. 

\subsection{Convergence}
Pulses with $N_\mathrm{seg} \ge 24$ lead to reliable convergence in several calibration runs, typically within 15 iterations as shown in \reffig{fig:main}c. Thus, all experiments were performed using \SI{36}{ns} long gates. An exception is the extraction of gate trajectories in \reffig{fig:main}d-e, where \SI{24}{ns} long gates were used. We have chosen the calibration algorithm such that it only adjusts those segments $\eps_j^g$ which are not at the baseline, resulting in 50 free parameters for the \SI{36}{ns} long $\pi/2_x$ and $\pi/2_y$ gates shown in \reffig{fig:main}\,a. Sometimes, moderate charge rearrangements in our sample lead to a deterioration of the optimized gates. As a remedy we run GSC again, resulting in slightly different gates than before. While the initial tune-up took about two days, recalibration from previously tuned-up gates only takes on the order of minutes to hours, depending on how much the sample tuning changed in between. To visualize the experimental gates, we perform self-consistent quantum state tomography (QST) \cite{Takahashi2013} and extract state information after each segment $\eps_j^g$. As seen in \reffig{fig:main}d-e, the qubit state trajectories for model and experiment closely resemble each other, indicating that the GSC-tuned pulses remain close to the optimum found in simulations.

\subsection{Fidelity and leakage benchmarking}
In order to determine the gate fidelity \fid, we apply randomized benchmarking (RB) after completion of GSC. In RB, the fidelity is obtained by applying sequences of randomly chosen Clifford gates, here composed of $\pi/2_x$ and $\pi/2_y$ gates, to the initial state \qZ. The last Clifford operation of each sequence is chosen such that \qZ would be recovered if the gates were perfect \cite{Magesan2011}. For imperfect gates, the return probability $p(\qZ)$ decays as a function of sequence length and the decay rate indicates the average error per gate. In addition, we also apply an extended RB protocol which omits the last Clifford from each RB sequence \cite{Wallman2015}. Without leakage, averaging over many randomly chosen sequences should yield $p(\qZ) =  \SI{50}{\%}$. However, for nonzero leakage we expect a single exponential decay of $p(\qZ)$ as a function of increasing sequence length since the additional leakage states have the same readout signature as \qO (see methods). 

\begin{figure}
	\centering
	\includegraphics[trim=0.0cm 0.0cm 0cm 0, clip,width=1\columnwidth]{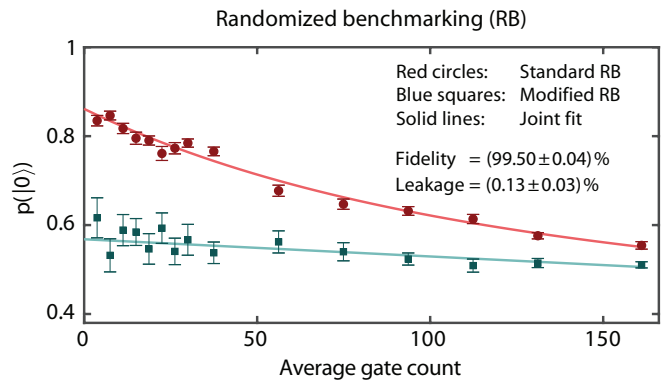}
	\caption{\textbf{Characterization of optimized gate sets.} The overall fidelity of a gate set consisting of \SI{36}{ns} long $\pi/2_x$ and $\pi/2_y$ gates is determined using RB (red) after a specific calibration run. Each data point is an average over 50 randomly chosen sequences of the respective length, the error bars indicate the standard error of the mean. In order to determine incoherent gate leakage we supplement the standard protocol (red) by a variant which omits the last inversion pulse (blue) \cite{Wallman2015}. Simultaneously fitting both curves yields $\fid = \SI{99.50 \pm 0.04}{\%}$ and a leakage rate $\leak = \SI{0.13 \pm 0.03}{\%}$. For this measurement, \dbz was stabilized with a standard deviation of $\sigma_{\dbz} = 2 \pi \cdot \SI{2.8}{MHz}$. The exact fit model is given in Supplementary Note 18.
	}
\label{fig:rb}	
\end{figure}

We indeed find such a decay law, indicated in blue in \reffig{fig:rb}. A joint fit of the standard (red) and leakage detection (blue) RB data yields $\fid = \SI{99.50+-0.04}{\%}$ and an incoherent gate leakage rate $\leak = \SI{0.13+-0.03}{\%}$ \cite{Wallman2015} (in the same GSC run we observed fidelities of \SI{99.4}{\%} and higher six times after different iterations in separate RB runs). Since our pulses operate close to the $\mathrm{S}-\mathrm{T_+}$ transition while $\ket{\mathrm{T_-}}$ is far away in energy, leakage should predominantly occur into the $\ket{\mathrm{T_+}}$ level. In another sample we observed \SI{0.4 \pm 0.1}{\%} leakage and much higher fast charge noise levels (indicated by a low echo time of $\Tte = \SI{183}{ns}$ for exchange oscillations) \cite{Cerfontaine2016}. This suggests that leakage is predominantly driven by charge noise.

\subsection{Agreement with theoretical results}
We previously predicted fidelities approaching \SI{99.9}{\%} for GaAs-based \sts qubits \cite{Cerfontaine2014} with the best reported noise levels \cite{Dial2013, Bluhm2010}. To make a sample-specific comparison, we measured \Tts and \Tte by performing free induction decay and echo experiments for both hyperfine and exchange driven oscillations at several detunings. For $\je = 2 \pi \cdot\SI{119}{MHz}$ we find that 726 coherent exchange oscillations are possible within $\Tte = \SI{6.1}{\micro s}$. This is larger than 585 oscillations reported in \citenumber{Dial2013} for a comparable charge noise sensitivity $dJ/d \eps \approx 2 \pi \cdot \SI{160}{MHz/mV}$. Based on these measurements, we construct a noise model including white charge noise and contributions from quasistatic hyperfine and charge noise (see Supplementary Note 3) to evaluate the fidelities of the numerically optimized gates used as a starting point for GSC (see Supplementary Note 6). Without systematic errors, their theoretical fidelities of \SI{99.86}{\%} ($\pi/2_x$) and \SI{99.38}{\%} ($\pi/2_y$) correspond to an average fidelity of \SI{99.62}{\%}. The good agreement with the experimentally observed average fidelity of $\SI{99.50+-0.04}{\%}$, which includes residual systematic errors, indicates that this noise model can be used to obtain good estimates of the achievable fidelities. Using this model we estimate that speeding up the gate pulses by a factor 6 would increase the fidelity at least to \SI{99.8}{\%} by further reducing the effect of \dbz fluctuations. Additional improvements are possible by using barrier control \cite{Martins2015} or reducing charge noise and residual systematic errors.

\section{Discussion}
One implication of our results is that the unavoidable presence of nuclear spins in GaAs spin qubits, which is often thought of as prohibitive for their technological prospects, actually does not preclude the fidelities required for fault-tolerant quantum computing. While the overhead associated with mitigating nuclear-spin-induced decoherence remains a disadvantage \cite{Bluhm2010}, GaAs quantum dot devices currently tend to be more reproducible and require less challenging lithographic feature sizes. Furthermore, they avoid the complication of several near-degenerate conduction band valleys and are better suited for the conversion between spin states and flying photonic qubits due to the direct band gap \cite{Kim2016, Joecker2018}. Given that GaAs has certain advantages over Si (but also other disadvantages), our study contributes to a well-founded comparison of the two material systems. While the measured infidelities are somewhat lower than achieved in some Si devices, further improvement is possible. Moreover, two-qubit gates will be a more decisive factor for scalability. Our simulations based on the same type of model validated by the present experiments predict two-qubit fidelities of \SI{99.9}{\%} \cite{Cerfontaine2019}.

Independent of the host material, a strength of \sts qubits are gate durations of a few tens of nanoseconds. These gates require only sub-gigahertz electrical control, in contrast to the manipulation frequencies typically used for single-spin qubits which often lead to slower gate speeds on the order of one microsecond. Furthermore, these relaxed control hardware requirements could substantially facilitate the adoption of integrated cryogenic control electronics. 

Although driven by the needs of GaAs-based \sts qubits, we expect that our approach is equally viable for other qubit types - specifically if many free gate parameters need to be tuned, if it is not intuitive how these parameters affect the qubit operations or if the qubit control model is not very accurate. As such, future work will focus on implementing exchange-mediated two-qubit gates \cite{Cerfontaine2019} in GaAs or Si, and experimentally studying the factors determining the success rate of GSC. With appropriate parallelization and simplifications (e.g. fixing the gradients used by GSC), our method could also be adapted for the calibration of larger qubit arrays.

\section{Methods}
\subsection{Qubit system}
This work was performed using two different samples. The first sample is identical to the one in \citenumber{Cerfontaine2016} and was used to establish the calibration routine and measure the Bloch sphere trajectories. The gate fidelities were obtained in a second sample with lower charge noise. 

We work in a dilution refrigerator at an electron temperature of about \SI{130}{mK}. A lateral double quantum dot is defined in the two-dimensional electron gas of a doped, molecular-beam epitaxy-grown GaAs/AlGaAs-heterostructure by applying voltages to metallic surface gates. For both samples, we use the same gate layout as \citenumber{Shulman2012} and \citenumber{Botzem2015} shown in \reffig{fig:main}b with two dedicated RF gates (yellow) for controlling the detuning. As we only apply RF pulses to these gates and no DC bias, we can perform all qubit operations without the need for bias tees, which reduces pulse distortions.

Quantum gates are performed in the (1,1) charge configuration, where one electron is confined in the left and one in the right quantum dot. In this regime, the computational subspace is defined by the $s_z = 0$ triplet state \qT and the spin singlet state \qS. The other $s_z = \pm 1$ (1,1) triplet states $\ket{\uparrow\uparrow}$ and $\ket{\downarrow\downarrow}$ are split off energetically via the Zeeman effect by applying an external magnetic field of \SI{500}{mT}.

We always read out and initialize the dot in the $(\qUD, \qDU)$ basis by pulsing slowly from (0,2) to (1,1) and thus adiabatically mapping singlet \qS and triplet \qT to \qUD and \qDU (see Supplementary Note 7).

\subsection{Readout calibration}
For measuring the quantum state, we discriminate between singlet and triplet states by Pauli spin blockade. Using spin to charge conversion \cite{Petta2005}, the resistance of an adjacent sensing dot depends on the spin state and can be determined by RF-reflectometry \cite{Reilly2007}. In this manner, we obtain different readout voltages for singlet and triplet states but cannot distinguish between $\ket{\mathrm{T}_0}$ and the triplet states $\ket{\mathrm{T}_{\pm}}$.

The measured voltages are processed in two ways. First, binning on the order of $10^4$ consecutive single shot measurements yields bimodal histograms where the two peak voltages roughly correspond to the singlet and triplet state. Second, the measured voltages are averaged over many repetitions of a pulse to reduce noise.

For the benchmarking experiments (which were performed using the second sample), we linearly convert the averaged voltages to probabilities $p(\ket{0})$. The parameters of the linear transformation are obtained by fitting the histograms with a model that takes the decay of \qT to \qS during the readout phase into account \cite{Barthel2009}. 

In the first sample we also observed considerable excitation of \qS to \qT. Thus, we modify the histogram fit model for the self-consistent state tomography data (which was obtained with the first sample) by introducing the excitation of \qS to \qT as an additional fit parameter.

For GSC, the averaged voltages $U_i$ corresponding to the error syndromes $S_i$ do not need to be explicitly converted to probabilities $p(\ket{0})$. Since mixed and triplet state reference voltages $U_{\mathrm{M}}$ and $U_{\mathrm{T}}$ are measured alongside the error syndromes, it is attractive to directly minimize the norm of $\tilde{U_i} = U_i - U_{\mathrm{M}}$ for $i = 1 \ldots 6$ and $\tilde{U_i} = U_i - U_{\mathrm{T}}$ for $i \in \{7, 8 \}$. 

Note that RB is insensitive to state preparation and measurement (SPAM) errors. In addition, we decrease the sensitivity of GSC to SPAM errors by using reference measurements for a completely mixed and a triplet state. Therefore, our readout calibration does not need to be especially accurate or precise.

Further information regarding readout can be found in Supplementary Notes 7 to 11.

\subsection{Data availability}
The benchmarking data that supports the key findings of this study are available in the supplementary information files. Various other raw data is available from the corresponding author upon reasonable request.

\subsection{Code availability}
The code used for this study is available from the corresponding author upon reasonable request. 

\subsection{Acknowledgments}
This work was supported by the Alfried Krupp von Bohlen und Halbach Foundation, DFG grants BL 1197/2-1, BL 1197/4-1, BO 3140/4-1 and SFB 1277. P.C. acknowledges support by Deutsche Telekom Stiftung. We thank Robert P. G. McNeil for fabricating the gate structures and ohmics of the second sample.

\subsection{Author contributions}
The first sample's heterostructure was grown with molecular-beam-epitaxy by D.B. and D.S. The second sample's heterostructure was grown by J.R., A.L. and A.D.W.; T.B. fabricated the first sample and set up the experiment. S.S.H. developed the driver for the digitizer hardware used for data acquisition. T.B. and P.C. developed the feedback software and conducted the experiment. H.B., T.B. and P.C. analyzed the data and co-wrote the paper. H.B. and P.C. developed the theoretical models.

\subsection{Competing interests}
The authors declare no competing interests.

\subsection{Additional information}
Correspondence and requests for materials should be addressed to P.C. or H.B. Online supplementary information accompanies this paper.

\end{document}